\documentclass[sigconf]{acmart}
\acmConference[arXiv]{}{2024}{US}

\usepackage{amsmath,amsfonts}
\usepackage{algorithm}
\usepackage{algorithmicx}
\usepackage{algpseudocode}
\usepackage{graphicx}
\usepackage{textcomp}
\usepackage{listings}
\usepackage{xcolor}
\usepackage[english]{babel}
\usepackage{color}
\usepackage{xspace}
\usepackage{multirow}
\usepackage{graphicx}
\usepackage{booktabs}
\usepackage{url}
\usepackage{xurl}
\usepackage{tipa}
\usepackage[T1]{fontenc}
\usepackage[utf8]{inputenc}
\usepackage{threeparttable}
\usepackage{hyperref}
\hypersetup{
	pdftitle={},
	pdfauthor={},
	pdfsubject={},
	pdfkeywords={},
	bookmarksnumbered=true,
	bookmarksopen=false,
	colorlinks=true,
	pdfstartview=FitB,
	pdfpagemode=UseOutlines
}

\definecolor{mygreen}{rgb}{0,0.6,0}
\definecolor{mygray}{rgb}{0.5,0.5,0.5}
\definecolor{mymauve}{rgb}{0.58,0,0.82}

\renewcommand\footnotetextcopyrightpermission[1]{} % removes footnote with conference information in first column
\newcommand{\ie}{\textit{i.e.,}\xspace}    % in other words
\newcommand{\eg}{\textit{e.g.,}\xspace}    % for example

\newcommand{\squishlist}{
	\begin{list}{$\bullet$} {
			\setlength{\itemsep}{0pt}
			\setlength{\parsep}{0pt}
			\setlength{\topsep}{0pt}
			\setlength{\partopsep}{0pt}
			\setlength{\leftmargin}{1.0em}
			\setlength{\labelwidth}{1em}
			\setlength{\labelsep}{0.5em}
		}
}
\newcommand{\squishend}{
	\end{list}
}

\newcommand{\code}[1]{\texttt{#1}}
\newcommand{\tool}{\text{E\&V}\xspace}

\newcommand{\cut}[1] {}

\begin{document}

\title{\tool: Prompting Large Language Models to Perform Static Analysis by Pseudo-code Execution and Verification}

\author{Yu Hao}
\orcid{0000-0002-3944-3162}
\affiliation{%
	\institution{UC Riverside}
	\city{Riverside}
	\country{USA}
}
\email{yhao016@ucr.edu}

\author{Weiteng Chen}
\affiliation{%
	\institution{Microsfot Research}
	\city{Redmond}
	\country{USA}
}
\email{weitengchen@microsoft.com}

\author{Ziqiao Zhou}
\affiliation{%
	\institution{Microsfot Research}
	\city{Redmond}
	\country{USA}
}
\email{ziqiaozhou@microsoft.com}

\author{Weidong Cui}
\affiliation{%
	\institution{Microsfot Research}
	\city{Redmond}
	\country{USA}
}
\email{Weidong.Cui@microsoft.com}

\begin{abstract}
Static analysis, the process of examining code without executing it, is crucial for identifying software issues. 
Yet, static analysis is hampered by its complexity and the need for customization for different targets. 
Traditional static analysis tools require extensive human effort and are often limited to specific target programs and programming languages. 
Recent advancements in Large Language Models (LLMs), such as GPT-4 and Llama, offer new capabilities for software engineering tasks. 
However, their application in static analysis, especially in understanding complex code structures, remains under-explored. 

This paper introduces a novel approach named \tool, which leverages LLMs to perform static analysis. 
Specifically, \tool employs LLMs to simulate the execution of pseudo-code, effectively conducting static analysis encoded in the pseudo-code with minimal human effort, thereby improving the accuracy of results. 
\tool includes a verification process for pseudo-code execution without needing an external oracle. 
This process allows \tool to mitigate hallucinations of LLMs and enhance the accuracy of static analysis results. 
We have implemented \tool in a prototype tool designed for triaging crashes through backward taint analysis. 
This prototype, paired with GPT-4-32k, has been applied to triage 170 recently fixed Linux kernel bugs across seven bug categories. 
Our experiments demonstrate that the prototype correctly identifies the blamed function in 81.2\% of the cases. 
Additionally, we observe that our novel verification process significantly improves the accuracy, increasing it from 28.2\% to 81.2\%.
\end{abstract}

\maketitle

\section{Introduction}
Static analysis plays a vital role in identifying potential issues early, thus improve software quality. 
However, developing tools for static analysis is challenging due to its complexity and the depth of understanding required, thus necessitating specialized expertise in static analysis.
Besides, developing such tools also involves significant engineering work. 
For instance, RETracer~\cite{DBLP:conf/icse/CuiPCFK16}, a static analysis tool that mainly performs backward taint analysis for triaging crashes in Windows systems, consists of over 66,000 lines of code in C and C++. 
This illustrates the extensive human effort and specialized knowledge needed. 
Additionally, these tools are often designed for specific target programs. 
If one wishes to apply the same static analysis to different programs or even different programming languages, the tool must be adapted or redeveloped for each new target. 
This limitation means that static analysis tools are not universally applicable without significant modifications.

Large Language Models (LLMs), such as GPT-4~\cite{DBLP:journals/corr/abs-2303-08774}, Llama 2~\cite{DBLP:journals/corr/abs-2307-09288} and Codex~\cite{DBLP:journals/corr/abs-2107-03374}, have shown potential in various software engineering tasks~\cite{DBLP:journals/corr/abs-2305-12138}. 
For example, they can assist in writing and reviewing code, detecting basic bugs, and creating code documentation~\cite{DBLP:journals/corr/abs-2308-10620}. 
However, the usage of LLMs in the static analysis, especially in the complex one, \eg inter-procedural static analysis, which tries to understand the relationships and interactions between different functions of a program, is not yet fully explored.
These models often struggle with understanding complex code structures.
They may also misinterpret code or produce inaccurate results, a problem known as hallucinations of LLMs~\cite{DBLP:journals/corr/abs-2309-01219}.
Recently, people have tried to resolve this by self-verification~\cite{weng2023large}, \ie LLMs check the results by themselves.
However, whether this approach is effective or not still needs to be explored~\cite{DBLP:journals/corr/abs-2206-10498, valmeekam2023large, DBLP:journals/corr/abs-2310-01798}.

Our research aims to enhance the capabilities of LLMs in performing static analysis tasks. 
We intend to design methods that allow these models to understand and analyze code more accurately, aligning with the specific analysis goals set by users. 
We also hope to reduce the need for extensive human effort when developing static analysis tools and minimize the occurrence of incorrect results caused by hallucinations of LLMs. 
By doing so, our work seeks to bridge the gap between the current capabilities of LLMs and the practical needs of static analysis in software engineering. 
This has the potential to significantly streamline the process of identifying and fixing code issues, reducing reliance on experts. 

% \ziqiao{Verification -> Validation? I think it may cause some misleading to
% readers. I thought we only have a set of checks we did for the GPT
% returned result. I propose to use "Result validation"?}
% \yu{I am not sure about this. And based on the info from GPT, it seems that Validation includes verification and some more things outside the scope of our method. So I think we can use verification here.}
In this paper, we present a novel method named \tool (\ie Execution and Verification) for conducting static analysis using LLMs. 
First, we need the human to construct the pseudo-code for the specific static analysis needed. 
Pseudo-code, a simplified version of programming code, is designed for easy comprehension by humans.
And LLMs have also demonstrated their ability to understand the pseudo-code\cite{DBLP:journals/corr/abs-2303-12712}. 
Besides, the development of pseudo-code requires less effort compared to directly developing static analysis tools. 
Next, we prompt LLMs to simulate or emulate the execution of this pseudo-code. 
By doing this, LLMs can effectively carry out the static analysis as if the code were being run in a real environment. 
Since LLMs can understand different programming languages, they can directly perform analysis based on the same pseudo-code for different programming languages.
This eliminates the need to develop different static analysis tools for different programming languages.
In addition, we propose a general and lightweight approach to verify the execution of the pseudo-code, reducing inaccuracies of the execution without the need for external verification oracles~(see \S\ref{sec:Verification} for more details).
By verifying the execution of pseudo-code, \tool indirectly mitigates the hallucinations and improves the accuracy of the static analysis results.

We have applied \tool in a prototype specifically designed for triaging crashes for the Linux kernel with GPT-4-32k, which mainly performs a backward taint analysis. 
Our evaluation of this prototype covered seven bug categories, including a total of 170 real recent crashes from the Linux kernel. 
The results are promising: the prototype achieved an overall accuracy rate of 81.18\%. 
Further, we conducted a thorough manual analysis of the results. 
This detailed review revealed that our prototype is effective in correctly performing backward taint analysis. 
Additionally, we have identified and summarized the factors contributing to the cases where our prototype did not yield the correct results. 
These insights are crucial for refining our prototype further. 
A significant breakthrough in \tool is the verification for the execution of pseudo-code. 
This novel verification process has greatly improved the accuracy rate of our prototype, elevating it from 28.24\% to 81.18\%. 
We also go into the details of verification for the execution, illustrating how this verification process can indirectly improve the accuracy and reliability of LLMs-based static analysis results. 
We summarize our main contributions as below:
\squishlist
    \item \textbf{New Method for Static Analysis:}
    We introduce a novel approach \tool that prompts LLMs for static analysis tasks. 
    By leveraging pseudo-code, we prompt LLMs to simulate the execution of static analysis processes. 
    This method represents a shift in how static analysis can be conducted by LLMs, reducing the reliance on extensive human effort and technical expertise typically required for developing static analysis tools.
    \item \textbf{Prototype Implementation and Open Sourcing:} 
    We have applied \tool in a prototype, specifically designed for triaging crashes of the Linux kernel through backward taint analysis. 
    The prototype demonstrates the practical application of our method. 
    To foster further research in this area, we will make our prototype and results open source, allowing others to access, use, and improve upon our work.
    \item \textbf{Comprehensive Evaluation:} 
    We perform a comprehensive evaluation of the prototype using real-world scenarios, including 170 recent crashes from the Linux kernel across seven bug categories. 
    We achieved an accuracy rate of 81.18\%, showcasing the effectiveness of \tool in real-world applications. 
    This novel verification process has greatly improved the accuracy rate of our prototype, elevating it from 28.24\% to 81.18\%. 
    Additionally, our in-depth analysis of the verification for the execution of pseudo-code offers valuable insights for future research about the mitigation of hallucinations.
\squishend

% \vspace{8pt}
% The rest of this paper is organized as follows: \S\ref{sec: Related Work} talks about related work, \S\ref{sec:Design} describes the overall design of our method, \S\ref{sec: triaging crashes} presents how we apply our method for triaging crashes, \S\ref{sec: Implementation} discusses its implementation, \S\ref{sec:evaluation} presents the evaluation results, \S\ref{sec: Limitations and Future Work} addresses limitations and future work, and \S\ref{sec: Conclusion} is conclusion.
\section{Related Work}
\label{sec: Related Work}

\subsection{Large Language Models for Program Analysis}

A lot of research has been conducted on the application of large language models (LLMs) in software engineering~\cite{hou2023large, fan2023large, wang2023software}, with a primary focus on areas such as code generation, code summarization~\cite{liu2023code}, test generation~\cite{sun2023automatic, ahmed2023improving}, and patch generation~\cite{siddiq2023empirical, xia2023conversation}.
In addition to these applications, some studies have explored the use of LLMs for program analysis. 
For instance, the work presented in~\cite{fu2023chatgpt} provides a comprehensive analysis of ChatGPT's direct application in the detection of vulnerabilities, identification of vulnerability types, estimation of severity levels, and recommendation of patches.
LLift~\cite{li2023hitchhikers} integrates LLMs with static analysis tools to identify use-before-initialization bugs within the Linux kernel.
It does this by having the LLMs analyze post conditions for the functions, which are then utilized to find the bugs more accurately. 
Furthermore, the research in~\cite{chakraborty2023ranking} proposes to re-rank the LLM-generated loop invariants to reduce the cost of wasted verification effort.
Compared with these studies, \tool is more general because it is easily applicable to any static analysis tasks.
And \tool is carefully designed to mitigate the hallucinations of LLMs, including a novel verification process.

\subsection{Hallucinations in Large Language Models}

The survey by Zhang et al.\cite{zhang2023sirens} categorizes hallucinations in large language models (LLMs) inference into three distinct types:
(1) input-conflicting hallucinations, where LLM-generated content is at odds with user-provided input;
(2) context-conflicting hallucinations, which arise when the content contradicts the model's own previously generated output;
and (3) fact-conflicting hallucinations, which occur when the output is inconsistent with established real-world knowledge.
To mitigate these hallucinations, several strategies have been proposed.
For instance, decoding strategies aim to reduce hallucinations by carefully selecting output tokens based on the probability distributions computed by LLMs~\cite{DBLP:journals/information/ZarriessVS21}. 
This approach, however, is constrained when LLMs are accessed via limited APIs that do not expose token-level output probabilities, a common limitation with proprietary models.
Another method involves enhancing LLM outputs with external knowledge sources or using specialized tools for task-specific improvements~\cite{ren2023investigating, mialon2023augmented}.
This, too, presents challenges, particularly in program analysis tasks, which often require significant human effort to obtain relevant external knowledge or to develop such tools.
A third approach involves internal consistency checks within LLM outputs~\cite{huang2023survey}.
An example of this is the self-checking mechanism introduced by Manakul et al.~\cite{manakul2023selfcheckgpt}, which enables LLMs to assess their own consistency and thus mitigate hallucination.
However, the effectiveness of such self-correction mechanisms remains uncertain, as highlighted by recent studies~\cite{DBLP:journals/corr/abs-2206-10498, valmeekam2023large, DBLP:journals/corr/abs-2310-01798, stechly2023gpt4}.
Despite these efforts, the mitigation of hallucinations in LLMs is still an open problem.

To address the various hallucinations exhibited by LLMs, we propose \tool, which incorporates specific features to counter each type of hallucination.
For input-conflicting hallucination, a specialized verification component in \tool is specifically tailored to address this issue~(more details in \S\ref{sec:Verification}).
For context-conflicting hallucination, \tool avoids it because the intermediate results are organized using uniform JSON format instead of chat history~(more details in \S\ref{sec: Context Conflicting Hallucinations}).
For fact-conflicting hallucination, we follow the idea of Retrieval Augmented Generation~\cite{lewis2021retrievalaugmented}, which is to retrieve relevant/supporting documents for the generation of text generator models.
We first retrieve relevant source code.
We then force LLMs to execute the provided pseudo-code on the provided source code to mitigate this kind of hallucination~(more details in \S\ref{sec: Fact Conflicting Hallucinations}).

\subsection{Large Language Models Based Agents}
% \ziqiao{My feeling of this section is that Auto agent is promising, but we do
% not use that because two papers said they are not good. I think it is okay to say
% there are many different agents, some do task scheduling based on GPT's
% feedback, some plans based on deterministic strategies. We did not
% compare them and cannot say which is better. We only need to say we use a LLM
% agent with hardcoded strategies in our prototype since the strategy of our
% prototype is relative clear. We did not explore whether
% the stragegy is similar or different (I guess it could be different for different
% bug types) if we let it to work on other bug types.}

Building agents with LLMs as its core controller is a promising direction~\cite{weng2023prompt}.
For example, AutoGPT~\cite{AutoGPT} claims that it is the agent for everything.
GPT Engineer~\cite{gptengineer} is the agent that automatically develops software.
LLMs-based agents can exhibit reasoning and planning abilities comparable to traditional AI agents~\cite{xi2023rise}.
The survey~\cite{xi2023rise} summarizes server reasons why LLMs are suitable for agents.
First, LLMs demonstrate autonomy, demonstrating the ability to operate without direct intervention from humans
or others.
Second, LLMs are reactivity, which means that the agent can sense changes in its inputs and take appropriate actions.
Thirdly, LLMs are proactive, which not only react to their input but also engage in goal-oriented actions, demonstrating skills in reasoning and planning. 
Finally, LLMs show their social capabilities, allowing them to communicate and collaborate effectively with other agents and humans, thereby playing a vital role in social interactions and task execution. 
Together, these attributes make the LLMs ideal for AI agents and it is a promising future research direction.
Recently, LLM OS has become a new exciting research direction~\cite{LLMOS}.
People thought that in a few years, the LLM itself would be the CPU, and the context window of the LLM would be the RAM, like an operating system.
The LLM OS can utilize a browser, calculator, code interpreter, and other tools to complete tasks.
On the other hand, some works~\cite{DBLP:journals/corr/abs-2206-10498, valmeekam2023large} show that current LLMs actually can not plan.
Therefore, this direction needs to be explored in depth to fully reveal its potential and impact.

In our scenario, since the strategies of \tool is relatively straightforward, we currently do not fully rely on LLMs as the core controller of our agent~(more details in \S\ref{sec:Overview}).
Instead, we hardcoded planning strategies for the key components and only used LLMs as the controller of the possible other tools in the future.
\section{\tool Design}
\label{sec:Design}

\tool is designed to enhance the capabilities of LLMs in performing static analysis with the need for extensive human effort and technical expertise typically required for developing static analysis tools.
In this section, we describe the design choice of \tool, and the details of the key components.
First, we present the overview of how we leverage large language models for static analysis by utilizing pseudo-code~(\S\ref{sec:Overview}).
Then we describe the details of components: 
% \textbf{Iterative Agent}~(\S\ref{sec:Iterative Agent}), 
\textbf{Pseudo-code Execution}~(\S\ref{sec:Pseudo Code Prompts}), which represents the E in \tool, and \textbf{Execution Specifications Verification}~(\S\ref{sec:Verification}), which represents the V in \tool.

\begin{figure*}[t]
\centering
\includegraphics[width=0.65\linewidth]{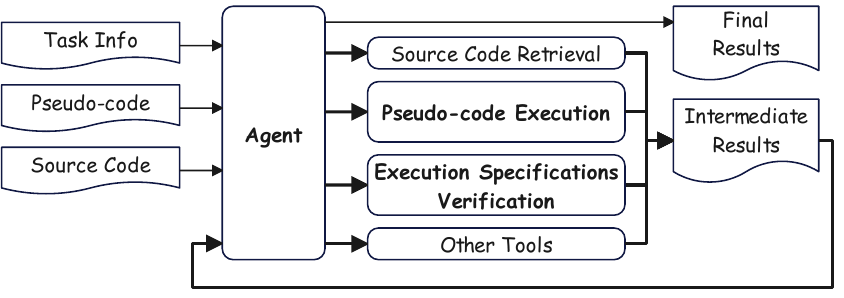}
\caption{Workflow}
\label{fig:Workflow}
\end{figure*}

\subsection{Overview}
\label{sec:Overview}
% The high level descriptions of each component

\autoref{fig:Workflow} presents the high-level workflow of \tool.
Firstly, \tool requires three types of input: 
(1) The task information, \eg triaging crashes and related crash reports. 
(2) The pseudo-code of the static analysis, which are used to guide the LLMs to perform the static analysis, and the execution specifications, which is used to verify the execution of the pseudo-code(more details in \S\ref{sec:Verification}).
(3) The source code of the target program to be analyzed.

% Briefly, given these inputs, the \textbf{Iterative Agent} would select appropriate components and construct the inputs for the components, based on both the inputs and the intermediate results obtained up to that point.
% The \textbf{Iterative Agent} would then send the inputs to the selected components, which would perform the corresponding analysis and return the intermediate results to the \textbf{Iterative Agent}.
% Then the \textbf{Iterative Agent} would parse the intermediate results and determine the next step of the analysis.
% This process is repeated until the final results are obtained.

% There are four components that can be selected by the \textbf{Iterative Agent}.
% The \textbf{Pseudo-code Execution} is used to perform each step of the static analysis by prompting LLMs to simulate the execution of the pseudo-code.
% This component is also the basis of the verification component.
% The \textbf{Source Code Retrieval} is used to extract the related source code, which is implemented by traditional static analysis in order to handle the extensive codebase.
% The \textbf{Execution Specifications Verification} is used to verify the execution of the pseudo-code to eliminate input-conflicting hallucinations~\cite{DBLP:journals/corr/abs-2309-01219} produced by the LLMs.
% Besides, the possible \textbf{Other Tools} can be integrated into \tool and would assist other analyses if necessary.

% \subsection{Iterative Agent}
% \label{sec:Iterative Agent}

The \textbf{Agent} plays a central role in \tool. 
It controls the entire workflow, manages the selection of appropriate components, constructs the inputs for other components, and organizes intermediate results.
In contrast to traditional static analysis approaches, which typically employ a single technique throughout the entire analysis process, \tool dynamically switches between various sub-components. 
This flexibility is used to take advantage of the different sub-components and efficiently manage various types of intermediate results that may arise during the execution of the workflow. 
Currently, we opt for hardcoded planning strategies rather than delegating them to LLMs due to the strategies of \tool is relatively straightforward.
% Currently, we opt for hardcoded planning strategies rather than delegating them to LLMs due to the absence of a definitive answer on whether LLMs are capable of planning~\cite{DBLP:journals/corr/abs-2206-10498, valmeekam2023large}. 

Specifically, the \textbf{Agent} utilizes the \textbf{Pseudo-code Execution} to perform a single step of the static analysis at a time, instead of conducting an end-to-end static analysis on all source code.
For instance, it may focus on a single function or variable, embodying the principle of task decomposition.
It constructs the inputs, \ie prompts, based on the task information, pseudo-code of static analysis, and intermediate results~(\eg previous analysis results, relevant source code, and verification results). 
Then the \textbf{Agent} would dispatch the prompts to the \textbf{Pseudo-code Execution}, to prompt LLMs to simulate the execution of the pseudo-code~(more details in \S\ref{sec:Pseudo Code Prompts}). 
After obtaining the corresponding responses from LLMs, the analysis results are transferred to a specialized verification component, referred to as \textbf{Execution Specifications Verification}~(more details in \S\ref{sec:Verification}). 
% Instead of directly verifying the analysis results of LLMs, this component is designed to indirectly verify the execution of pseudo-code to eliminate input-conflicting hallucinations produced by the LLMs and ensure that the iterative analysis proceeds. 
% \wc{this paragraph is about agent and we already added a reference here, do we still need to highlight it?}
In cases where the results do not pass the verification, the \textbf{Agent} initiates a re-analysis by re-constructing new prompts and requesting the \textbf{Pseudo-code Execution}, which then conducts a new round of analysis, focusing on the same sub-analysis task.
Besides, the \textbf{Source Code Retrieval} is used to extract the related source code, which is implemented by traditional static analysis in order to handle the extensive codebase.
And the possible \textbf{Other Tools} can be integrated into \tool and would assist other analyses if necessary.
% \wc{The whole paragraph overlaps a lot with the overview, we could merge it into overview.}

\subsubsection{Mitigation of Context-conflicting Hallucinations}
\label{sec: Context Conflicting Hallucinations}

Context-conflicting hallucinations arise when the content generated by LLMs is inconsistent with the previously generated outputs by models themselves.
A straightforward approach to organize the intermediate results is to use the chat history of LLMs, which inevitable contains the context of each inference of LLMs and brings the context-conflicting hallucinations.
Another disadvantage of using chat history is that the reliance on chat history could introduce a form of inertia, \ie LLMs tends to produce results that are similar to the previous results in the analysis process or format.
The chat history may overly influence the current output, leading to results that are not purely based on the current input but are also influenced by chat history, somewhat similar to few-shot learning\cite{parnami2022learning}. 

In fact, in the design of \tool for static analysis, only the results of chat history are necessary. 
The other parts, \eg thinking steps, and reasons for the results from chat history are not necessary.
In order to mitigate this kind of hallucinations, we choose to avoid the context of each inference of LLMs as much as possible.
So the \textbf{Agent} would not use the chat history of LLMs as the input of the next inference of LLMs.
Instead, the \textbf{Agent} would summarize the previous analysis and use uniform JSON format intermediate results.
The summarized JSON results can minimize the context by only containing the analysis results, without previous analysis process, reasons, oven even format in the chat history of LLMs.

Besides, in addition to mitigating the context-conflicting hallucinations, the uniform JSON format makes it significantly more compatible with various components in a system, especially those not based on LLMs, \eg \textbf{Source Code Retrieval}, \textbf{Execution Specifications Verification} and other possible tools.

\subsubsection{Mitigation of Fact-conflicting Hallucinations}
\label{sec: Fact Conflicting Hallucinations}

The fact conflicting hallucinations~\cite{zhang2023sirens} refer to the hallucinations that the content generated by LLMs conflicts with established world knowledge.
When performing the static analysis, the relevant source code is one of the most necessary fact, \eg function definition of the analyzed function.
However, due to the token limitations of LLMs, it is impossible to directly embed the entire source code base, \eg the Linux kernel, at once in the prompts. 
Retrieval Augmented Generation~\cite{lewis2021retrievalaugmented} is a promising direction to mitigate this kind of hallucination, which is to retrieve relevant/supporting documents for the generation of text generator models.
Therefore, the \textbf{Source Code Retrieval} is designed to extract the relevant source code, which is implemented by traditional static analysis in order to handle the extensive codebase~(more details in \S\ref{sec: Source Code Query}).

When constructing the prompts for \textbf{Pseudo-code Execution}, the \textbf{Agent} would first summarize the list of names of the required functions or structures, which can be obtained from the intermediate results, \eg the results of previous round of \textbf{Pseudo-code Execution} or the possible \textbf{Other Tools}.
Then the \textbf{Agent} would send the list of names to the \textbf{Source Code Retrieval} to retrieve the relevant source code. 
In this way, the \textbf{Pseudo-code Execution} can utilize the retrieved source code to mitigate the fact-conflicting hallucinations.
\begin{figure}[t]
    \centering
    \includegraphics[width=0.75\linewidth]{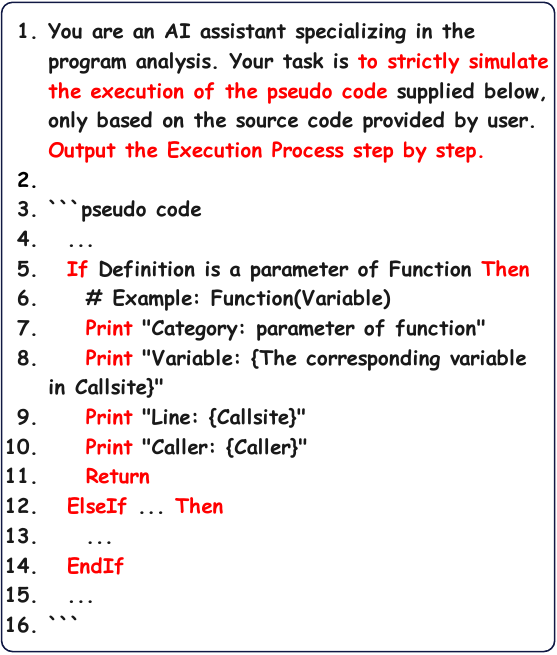}
    \caption{Example Prompt of Pseudo-code Execution}
    \label{fig:Pseudo Code Prompt Example}
\end{figure}

\subsection{Pseudo-code Execution}
\label{sec:Pseudo Code Prompts}

Employing LLMs for static analysis with straightforward prompts written in natural language yields unreliable and sometimes incorrect results. 
To mitigate this, we propose \textbf{Pseudo-code Execution} to enhance the reliability and precision of LLMs in performing static analysis. 
GPT-4 has demonstrated its capability in understanding pseudo code~\cite{DBLP:journals/corr/abs-2303-12712, DBLP:journals/corr/abs-2305-11790}. 
Given that static analysis can be precisely expressed in pseudo-code, it is intuitive that pseudo-code could serve as an effective medium for prompts of static analysis. 
As shown in \autoref{fig:Pseudo Code Prompt Example}, we first define the role of LLMs and the task is to let LLMs to \code{``strictly simulate the execution of the pseudo code''}. 
Following this, the pseudo-code for the static analysis, \eg backward taint analysis for triaging crashes, is provided, enabling LLMs to perform required static analysis and allowing users to have more precise control of LLMs. There are several advantages associated with \textbf{Pseudo-code Execution}:

\textbf{More Precise Prompts}. 
    Leveraging well-structured control flow in pseudo-code, we can design prompts with less ambiguity.
    This, in turn, facilitates more precise control over LLMs than natural language prompts. 
    The LLMs would be prompted to follow the sequences, branches, loops and function calls to take different actions.
    % For example, as shown in \autoref{fig:Pseudo Code Prompt Example}, 
    % LLMs would first consider the condition of the \code{If} statement at line 5. 
    % When the condition is satisfied, the LLMs would then proceed to the \code{Then} branch at line 6. 
    % Conversely, when the condition is not satisfied, the LLMs would proceed to the \code{ElseIf} branch at line 12. 
    % \wc{Is it necessary to explain how If-Then-Else works?}
    This level of precision is hard to achieve with natural language prompts, especially with limited tokens.

\textbf{Debugging Capability}. 
    When we design the pseudo-code of, it is sometimes difficult to figure out which parts of the prompts are causing wrong analysis results
    By integrating prompts that mandate LLMs to \code{``Output the Execution Process step by step''}, LLMs can output intermediate execution steps of the pseudo-code for developers to debug the execution and refine the pseudo-code, analogous to utilizing print statements for debugging in traditional programming.
    If any differences, which is between the design of the pseudo-code and the actual execution, occur in the intermediate outcomes, we can adjust the relevant logic statements (\eg an \code{If} or \code{ElseIf} condition) within the pseudo-code construct.
    This debugging capability is highly dependent on pseudo-code prompts.
    Unlike straightforward natural language, pseudo-code has a well-structured control flow, which allows us to debug the thinking process of LLMs based on the control flow graph.

\textbf{Automatic Verification Mechanisms}.
    Because of the control flow in pseudo-code and the LLMs are prompted to \code{``strictly simulate the execution of the pseudo code''}, the analysis results must be wrong if there are some violations during the execution.
    In other words, we can verify the execution of pseudo-code to infer whether the analysis results are wrong or not. 
    From this insight, we can utilize execution specifications from the pseudo-code to establish a framework for the automatic verification of LLMs analysis results. 
    More details are discussed in \S\ref{sec:Verification}.

\textbf{Finer Grained Task Decomposition}. 
    The strategy of task decomposition via prompt engineering is widely recognized as an effective method for enhancing the performance of LLMs, as evidenced in studies such as CoT~\cite{wei2023chainofthought} and ToT~\cite{yao2023tree}.
    To optimize the efficacy of LLMs, it is beneficial to either construct straightforward static analysis pseudo-code or to decompose complex static analysis tasks into a series of simpler ones. 
    This strategy facilitates easier processing and interpretation by LLMs, enhancing overall analysis quality.
    Unlike natural language prompts, which are sometimes ambiguous and difficult to decompose, the precision control provided by pseudo-code execution enables a finer grained task decomposition.
    can be designed to be more precise and unambiguous.
    Just like to refactor the code, we can refactor the pseudo-code to decompose the complex analysis task into a series of simpler ones.
    Additionally, given that LLMs have the capacity to help refactor the code~\cite{fan2023large}, it is plausible to utilize LLMs to refactor the pseudo-code as well
    (see more discussion on automatic task decomposition in \S\ref{sec: Limitations and Future Work}).

\textbf{Wide Range of Applications}.
    Given the advanced capabilities of LLMs to interpret and process various constructs in pseudo-code, such as sequences, branches, loops, and recursive functions, their application in static analysis shows significant promise. 
    This versatility allows for their broad application across diverse static analysis tasks.
    And all of the above advantages are also applicable to other static analysis tasks.

Although \textbf{Pseudo-code Execution} has many advantages, there is a limitation that the pseudo-code must be designed carefully with enough task decomposition.
Considering the capabilities of different LLMs, it is necessary to perform more task decompositions to ensure that LLMs can handle complex analysis, \eg inter-procedural analysis, loops, and recursive functions in the pseudo-code.

For instance, let us consider the scenario of inter-procedural analysis. 
This type of analysis involves the analysis of multiple functions in a program, such as the caller and callee functions. While LLMs are capable of performing such analysis, direct implementation of complex inter-procedural analysis in pseudo-code might not be ideal. 
This is particularly relevant when the source code for the caller or callee functions is not provided to the LLMs. 
In such cases, LLMs might proceed without requiring the source code, leading to potential hallucinations in the analysis. 
A more effective method would break this process. 
Initially, the LLM would focus on identifying and outputting the names of the caller or callee functions for further analysis. 
Subsequently, an external tool~(\eg \textbf{Source Code Retrieval} in \S\ref{sec: Source Code Query}) could be employed to supply the relevant source code for these identified functions. 
With this information at hand, the LLMs could then initiate a new round of more accurate and reliable analysis.

\subsection{Execution Specifications Verification}
\label{sec:Verification}

The input-conflicting hallucinations~\cite{zhang2023sirens} refer to the hallucinations that the content generated by LLMs conflicts with the input provided by users.
Specialized external tools like code executors and calculators have been employed to counter these hallucinations in various general-purpose applications~\cite{DBLP:journals/corr/abs-2305-11738, DBLP:journals/corr/abs-2307-13528}.
However, these methods are not directly applicable to the domain of static analysis. 
This is primarily because static analysis lacks an oracle for the clear validation of generated outputs. 
Even in rigorously designed static analysis tools, the challenges of false positives and false negatives remain unresolved.
On the other side, the work~\cite{manakul2023selfcheckgpt} attempts to mitigate these kinds of hallucinations by proposing a self-checking mechanism to detect consistency by LLMs themselves.
However, it is still unknown whether these mechanisms can really work or not~\cite{DBLP:journals/corr/abs-2206-10498, valmeekam2023large, DBLP:journals/corr/abs-2310-01798, stechly2023gpt4}.
In our scenario, we found that the self-checking mechanism is not reliable enough to detect the hallucinations produced by LLMs in static analysis tasks.

To address this issue, we introduce a novel methodology which is called \textbf{Execution Specifications Verification}. 
This methodology is specifically designed to mitigate input-conflicting hallucinations in LLMs when applied to static analysis tasks with \textbf{Pseudo-code Execution}. 
There are two results in the \textbf{Pseudo-code Execution}: the execution process based on the control flow of pseudo-code, and the analysis results based on the goal of the pseudo-code.
% \wc{what is execution of pseudo-code?}
Just like the real code, if the code is executed, the results of the code will naturally be generated.
When we prompt LLMs to simulate the execution of pseudo-code of static analysis, the results of static analysis will also be generated naturally.
% \wc{what is analysis results?}
The key insight of \textbf{Execution Specifications Verification} is that we can indirectly determine whether the analysis result is wrong or not by verifying the execution of pseudo-code. 
If the execution of pseudo-code is wrong, the analysis result must be wrong.

Specifically, outputs, which serve either debugging or verification purposes, could be generated by leveraging pseudo-code with programming-language-like constructs, such as the \code{print} function. 
In addition to the pseudo-code, a set of specifications for the execution of pseudo-code, which describes the correct execution of the pseudo-code, are also provided to the \tool.
Failure to meet these specifications triggers an immediate re-analysis by the LLMs. 
During the re-analysis, we would increase the temperature of the LLMs to encourage more random outputs. 
This process is repeated until the outputs meet the specifications. 

We classify violations of specifications into three categories and the contributions of those violations in \textbf{Execution Specifications Verification} are evaluated in \S\ref{sec:specification}:

\textbf{Unrecognized Execution.} 
    If the analysis is guided by the pseudo-code and produces explicitly well-defined outputs, it becomes crucial to predetermine an accepted set of categories for these outputs. 
    Any differences from this predetermined set are a kind of violation of the specifications. 
    For example, as shown in line 7 in \autoref{fig:Pseudo Code Prompt Example}, accepted output categories are specified when a taint sink is identified as a function parameter, \ie \code{Category: parameter of function}. 
    A subsequent verification process ensures that this set of accepted categories is met. 
    The recognized category of output is also necessary for the \textbf{Agent} to choose the next sub-module.
    If the output is not recognized, the \textbf{Agent} would not know how to handle the output and would not be able to choose the next sub-module.

\textbf{Incomplete Execution.} 
    Each analysis could potentially generate multiple outputs. 
    To determine their completeness, strict verification procedures are required. 
    If the output is missing any of the specified elements, the output is considered incomplete and therefore violates the specification.
    For instance, lines 7-10 in \autoref{fig:Pseudo Code Prompt Example} shows that a complete output should include the \code{Category}, \code{Variable}, \code{Line}, and \code{Caller} function when a taint sink is used as a parameter. 
    Completeness is verified by examining the presence of all these constituent elements. 
    And a complete result is also necessary for the \textbf{Agent} to construct the complete inputs for the next sub-module.
    If the output is incomplete, the \textbf{Agent} would miss some inputs for the next sub-module.

\textbf{Inconsistent Execution.} 
    Given that an analysis could produce multiple outputs, it is important to establish specifications for consistency among these outputs. 
    For example, lines 8-10 in \autoref{fig:Pseudo Code Prompt Example} \code{print} the \code{Variable}, which is the corresponding variable in callsite, the \code{Line}, which is the callsite, and the \code{Caller} function, which is the function calling current function. 
    These elements are subsequently cross-verification to ensure internal consistency, including checks on the occurrence of the variable within the line of code, the existence of the line within the provided source code, the inclusion of the calling function in the function call trace, or the presence of a function call at the callsite line.
\begin{figure*}[t]
    \centering
    \includegraphics[width=0.85\linewidth]{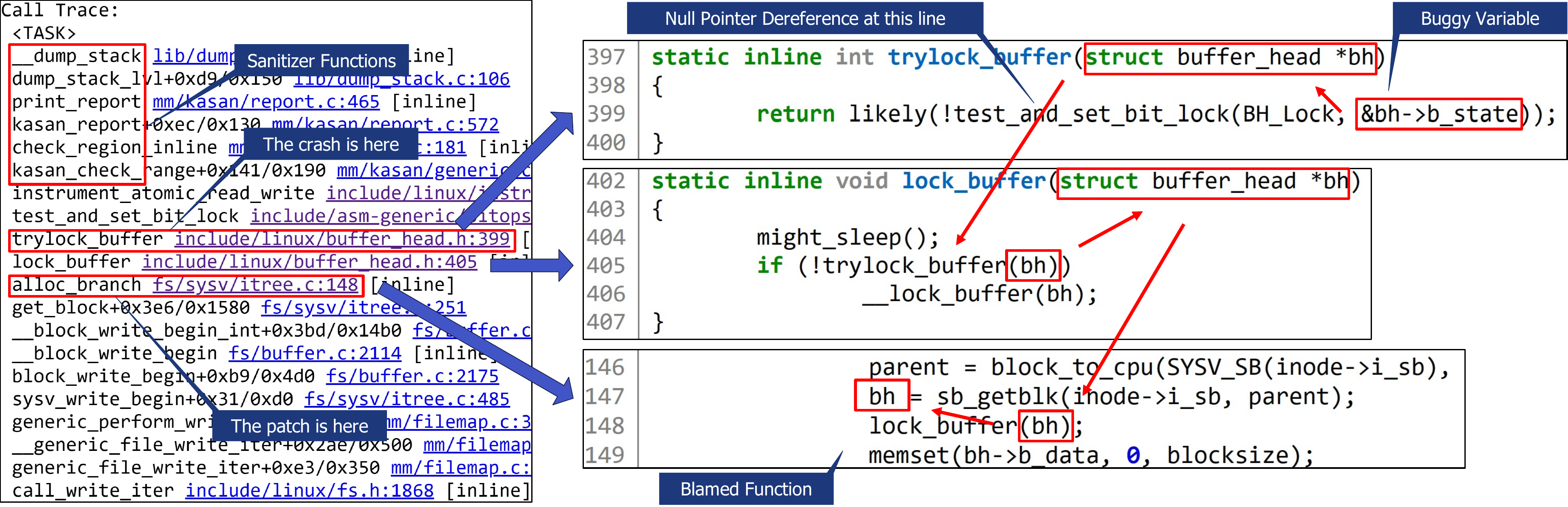}
    \caption{A Null Pointer Dereference Bug from Linux Kernel}
    \label{fig:Motivating Example}
\end{figure*}

\section{Triaging Crashes}
\label{sec: triaging crashes}

Triaging crashes in software is a critical process undertaken by many software providers, which involves identifying the root cause of a crash, grouping identical bugs, and assigning them to the respective developers responsible for the blamed functions. 
Traditional statistic based crash triaging methods, however, have limitations because they only rely on the syntactic information of stack traces at the time of the crash, without in-depth consideration of the semantics information of the program. 
% \wc{what are traditional methods? RETracer is not?}
Unlike traditional systems, RETracer~\cite{DBLP:conf/icse/CuiPCFK16} performs a backward taint analysis to find the blamed functions, where the patches should be located or nearby, offering a deeper understanding of how different functions on the stack contribute to a crash.

Triaging crashes is also a critical process in the Linux kernel community.
Syzbot~\cite{syzbot} is a continuous fuzzing system for the Linux kernel.
% \wc{we need a better transition.}
It has found and fixed more than 4,800 bugs in the Linux kernel. 
However, there are also more than 8,000 bugs found by syzbot that have not been fixed.
One reason is that these bugs were not even sent to the right developers~\cite{LPCSyzbot}. 
However, we could not directly apply RETracer to this problem without a lot of engineering efforts, because RETracer is specifically designed for Microsoft Windows and Office, which takes more than 66,000 lines of code in C and C++ to implement.
% \wc{So? It is difficult to port it?}
In order to resolve this problem, we applied \tool and developed a prototype following the same idea of RETrace. 
At the same time, we want to demonstrate the effectiveness of \tool in performing static analysis.

\subsection{A Motivating Example}
\label{sec: Motivating Example}
  
\autoref{fig:Motivating Example} illustrates a null pointer dereference bug originating from the Linux kernel and found by syzbot. 
Generally, because fuzzing detects bugs through the sanitizer, functions at the beginning of the call trace~(\eg \code{kasan\_report}) are sanitizer-instrumented functions. 
The actual null pointer dereference occurs in the \code{trylock\_buffer} function, specifically at line 399 where the variable \code{bh} is dereferenced.
% \wc{line 399 actually is not where pointer dereference happens due to the \& operator.}
To identify the blamed function, it is necessary to perform a backward trace on the variable \code{bh} based on the idea of RETracer. 
Since \code{bh} is passed as a parameter to \code{trylock\_buffer}, its origins must be traced recursively to the caller function. 
A similar tracing effort reveals that \code{bh}, located at line 405, is also passed as a parameter to the \code{lock\_buffer} function. 
Eventually, it is determined that \code{bh} is derived from the return value of a function call at line 147 in \code{alloc\_branch}, which is subsequently identified as the blamed function. 
The patch is also located in this function.

\subsection{Application}
\label{sec: Application in Triaging Crashes}

This section will present how to apply \tool to triaging crashes for bugs in the Linux kernel found by fuzzing.
First of all, there are three types of input data that are available to the Agent:
(1) The task information, which is crash triaging and crash reports, including the call trace obtained from fuzzing and any related debug information, \eg the call trace in \autoref{fig:Motivating Example}; 
(2) The pseudo-code of the static analysis, \ie the backward taint analysis, and the execution specifications; 
(3) The source code of the Linux kernel. 

With these inputs, the task of crash triaging mainly involves two basic sub-tasks: 
(1) Identification of the variable within the call trace responsible for the crash, \eg the variable \code{bh} in the motivating example; 
(2) Performing a backward taint analysis on this variable to trace back to the blamed function. 

% \ziqiao{No need to use \textbf{some terms} for each occurrence of those terms. This is a general comments the paper.}

For the first sub-task, the Agent consults the LLM by directly constructing straightforward prompts to identify the variable responsible for the crash based on the bug reports and related debug information.
Then the Agent would perform a backward taint analysis on the variable through Pseudo-code Execution component. 
The Source Code Retrieval component would be used to retrieve the related source code of the Linux kernel to mitigate the fact-conflicting hallucinations.
And the Execution Specifications Verification component would be used to verify the execution of the pseudo-code.
After several iterative cycles of backward taint analysis, the blamed function is identified.

In fact, the most straightforward prompt to perform a backward taint analysis is directly asking LLMs to do it. 
However, the results of those kinds of straightforward prompts are unstable and error-prone. 
% \ziqiao{I do not think LLM "understand" it. LLM only returns the correct definition of it and then can do something similar based on examples it saw from its training set.}
While LLMs know the overall concept of backward taint analysis, they lack the specific knowledge required for effective triaging crashes, such as the desired types of taint sources and sinks. 
One alternative is to encode that information directly into the natural language prompt used for guiding the backward taint analysis. 
However, it is hard to avoid ambiguity in natural language prompts when the analysis becomes complex.
% \wc{I thought we already explain pseduo code is better, why explain it again?}

Briefly, the backward taint analysis is to trace the origin of a variable. 
And the taint source and the taint sink should be defined in the pseudo-code. 
The initial taint source is the variable responsible for the crash, which is from the input of the pseudo-code. 
For the taint sink, we use the same four kinds of taint sinks defined in RETrace, which are the constant value, the global variable, the return value of the function call, or the field of a structure. 
For other situations when tracing backward, we perform another round of backward taint analysis on the variable at the assignment statement or the parameter at the callsite. 
For example, if the taint source is passed as a parameter, we will perform another round of backward taint analysis on the parameter at callsite in the caller function. 
As shown in figure \autoref{fig:Pseudo Code Prompt Example} at lines 5-11, a \code{If} statement is used to check whether the taint source is passed as a parameter. 
In the \code{If} statement, we print some necessary information to perform another round of backward taint analysis in the caller function, \eg the corresponding variable and the caller function name.
If the taint source is assigned from another variable, we will also perform another round of backward taint analysis on the variable at the assignment statement. 
Besides, we can also define some additional taint sinks for different bug types in the pseudo-code. 
For example, for stack out of bound, the taint sink should also include the stack variable.
% \wc{it is hard to follow this paragraph.}
\section{Implementation}
\label{sec: Implementation}

In total, our implementation of the prototype consists of 2,700 LoC in Python and 300 LoC in C++. To summarize the code distribution:
(1) 500 LoC in Python is to parse crash reports generated from fuzzing.
(2) 1,700 LoC in Python is to manage the entire workflow, which includes the Agent and the Execution Specifications Verification component.
(3) A combined 300 LoC in C++ and 500 LoC in Python are utilized for Source Code Retrieval operations.
In the following sections, we will delve into further details regarding the formatting of results produced by LLMs and source code retrieval operations.

\subsection{Formatting for Intermediate Results}
\label{sec: Format}

As mentioned in \S\ref{sec:Overview}, we need to ensure that the output of each component is in JSON format.
% \wc{Looks redundant to me.}
For the Source Code Retrieval component, the Execution Specifications Verification component, and possibly other future components based on traditional static analysis, this is straightforward, as the output of these components can be fully controlled and is already in JSON format.
For Pseudo-code Execution component, which is based on LLMs, a straightforward approach to enforce this standard is to include format-specific instructions within the query prompts themselves. 
For instance, one can state, \code{"The output must be the following JSON structure"}. 
While this prompt ensures the output aligns with the JSON format, it has limitations in terms of flexibility. 
In cases where different output categories require different JSON structures, this approach needs us to specify separate formats for each category directly within the query prompts. 
This makes the design of prompt more complex and increases the length of prompts, which may reduce the performance of LLMs. 
Moreover, this method is particularly unsuitable for Pseudo-code Execution component.
The pseudo-code mainly employs the \code{Print} function for result output.
To address these challenges, we introduce a supplementary LLM that is responsible for summarizing and formatting the output of other sub-modules. 
This strategy enables us to consistently produce outputs in JSON format, even when using Pseudo-code Execution component, thereby simplifying their integration with other modules.

\subsection{Source Code Retrieval}
\label{sec: Source Code Query}

As mentioned in \S\ref{sec:Overview}, in order to mitigate the fact-conflicting hallucinations, the Source Code Retrieval component is designed to address this challenge.
In the crash triaging scenario, it is necessary to retrieve the relevant source code of the Linux kernel in the commit version where the crash happened.
Currently, we only support the source code retrieval of the functions or structures because they are the most important information for the analysis.
After the Agent constructs a list of names of the functions or structures needed for the analysis, the Source Code Retrieval component is responsible for retrieving the source code based on the name from the list.
There are three steps in the Source Code Retrieval component. 
Firstly, a scalable, traditional LLVM~\cite{DBLP:conf/cgo/LattnerA04} based method is utilized to locate the files including the functions or structures from the list.
It mainly relies on the Clang tool to parse the source code of the Linux kernel and then builds an index based on debug information for each file.
Then, for performance reasons, only the located files are checked out to the commit version where the crash happened.
% \wc{it is more intuitive to state the problem first.}
Because the included header files are not checked out, those located files can not be fully compiled and complete debugging information can not be generated.
The Source Code Retrieval component only utilizes the basic Abstract Syntax Tree and string pattern matching to retrieve the source code of the functions or structures.
In this way, Source Code Retrieval component can handle most of the functions or structures.
However, there are still some corner cases that can not be retrieved by this method, which is one of the main reasons for inaccuracies of the prototype, as we mentioned in \S\ref{sec:accuracy}.

% Please add the following required packages to your document preamble:
% \usepackage{booktabs}
\begin{table*}[t]
  \centering
  \caption{Accuracy of Identified Blamed Functions}
  \label{tab:Accuracy}
  \begin{tabular}{@{}r|r|rrr|rr|rr@{}}
  \toprule
  \multicolumn{1}{c|}{Bug Category} &
  \multicolumn{1}{c|}{Total} &
  \multicolumn{1}{c}{Function} &
  \multicolumn{1}{c}{Callee} &
  \multicolumn{1}{c|}{Related} &
  \multicolumn{2}{c|}{Correct} &
  \multicolumn{2}{c}{Wrong} \\ \midrule
  stack-out-of-bounds  & 28  & 11 & 7  & 6  & 24  & 85.71\% & 4  & 14.29\% \\
  slab-out-of-bounds   & 34  & 10 & 9  & 8  & 27  & 79.41\% & 7  & 20.59\% \\
  global-out-of-bounds & 11  & 5  & 2  & 1  & 8   & 72.73\% & 3  & 27.27\% \\
  invalid-free         & 16  & 7  & 3  & 3  & 13  & 81.25\% & 3  & 18.75\% \\
  double-free          & 4   & 1  & 0  & 1  & 2   & 50.00\% & 2  & 50.00\% \\
  use-after-free       & 33  & 7  & 4  & 13 & 24  & 72.73\% & 9  & 27.27\% \\
  null-ptr-def         & 44  & 23 & 4  & 13 & 40  & 90.91\% & 4  & 9.09\%  \\ \midrule
  Sum                  & 170 & 64 & 29 & 45 & 138 & 81.18\% & 32 & 18.82\% \\ \bottomrule
  \end{tabular}
\end{table*}

% Please add the following required packages to your document preamble:
% \usepackage{booktabs}
\begin{table*}[t]
  \centering
  \caption{Improvement of Execution Specifications Verification}
  \label{tab:Improvement of ESV}
  \begin{tabular}{@{}r|r|rr|rr|rr@{}}
  \toprule
  \multicolumn{1}{c|}{Bug Category} & \multicolumn{1}{c|}{Total} & \multicolumn{2}{c|}{PE} & \multicolumn{2}{c|}{PE+ESV} & \multicolumn{2}{c}{$\Delta$ESV} \\ \midrule
  stack-out-of-bounds  & 28  & 5  & 17.86\% & 24  & 85.71\% & 19 & 67.86\% \\
  slab-out-of-bounds   & 34  & 8  & 23.53\% & 27  & 79.41\% & 19 & 55.88\% \\
  global-out-of-bounds & 11  & 2  & 18.18\% & 8   & 72.73\% & 6  & 54.55\% \\
  invalid-free         & 16  & 5  & 31.25\% & 13  & 81.25\% & 8  & 50.00\% \\
  double-free          & 4   & 0  & 0.00\%  & 2   & 50.00\% & 2  & 50.00\% \\
  use-after-free       & 33  & 9  & 27.27\% & 24  & 72.73\% & 15 & 45.45\% \\
  null-ptr-def         & 44  & 19 & 43.18\% & 40  & 90.91\% & 21 & 47.73\% \\ \midrule
  Sum                  & 170 & 48 & 28.24\% & 138 & 81.18\% & 90 & 52.94\% \\ \bottomrule
  \end{tabular}
\end{table*}

% Please add the following required packages to your document preamble:
% \usepackage{booktabs}
% \usepackage{multirow}
\begin{table*}[t]
\centering
\caption{Rate of Execution Violation}
\label{tab:Rate of ESV}
\begin{tabular}{@{}r|r|rrr|rrr|c@{}}
\toprule
\multicolumn{1}{c|}{\multirow{2}{*}{Bug Category}} &
  \multicolumn{1}{c|}{\multirow{2}{*}{$\Delta$ESV}} &
  \multicolumn{3}{c|}{Execution} &
  \multicolumn{3}{c|}{Execution Violation} &
  \multicolumn{1}{c}{\multirow{2}{*}{\begin{tabular}[c]{@{}c@{}}Execution Violation\\ /Execution\end{tabular}}} \\ \cmidrule(lr){3-5}\cmidrule(lr){6-8}
\multicolumn{1}{c|}{} &
  \multicolumn{1}{c|}{} &
  \multicolumn{1}{c}{Sum} &
  \multicolumn{1}{c}{Avg} &
  \multicolumn{1}{c|}{Max} &
  \multicolumn{1}{c}{Sum} &
  \multicolumn{1}{c}{Avg} &
  \multicolumn{1}{c|}{Max} &
  \multicolumn{1}{c}{} \\ \midrule
stack-out-of-bounds  & 19 & 108 & 5.68 & 10 & 30  & 1.58 & 3 & 27.78\% \\
slab-out-of-bounds   & 19 & 169 & 8.89 & 19 & 38  & 2.00 & 4 & 22.49\% \\
global-out-of-bounds & 6  & 45  & 7.50 & 14 & 9   & 1.50 & 3 & 20.00\% \\
invalid-free         & 8  & 38  & 4.75 & 9  & 11  & 1.38 & 2 & 28.95\% \\
double-free          & 2  & 8   & 4.00 & 5  & 3   & 1.50 & 2 & 37.50\% \\
use-after-free       & 15 & 72  & 4.80 & 12 & 24  & 1.60 & 3 & 33.33\% \\
null-ptr-def         & 21 & 106 & 5.05 & 11 & 24  & 1.14 & 1 & 22.64\% \\ \midrule
Sum                  & 90 & 546 & 6.07 & 19 & 139 & 1.54 & 4 & 25.46\% \\ \bottomrule
\end{tabular}
\end{table*}

% Please add the following required packages to your document preamble:
% \usepackage{booktabs}
\begin{table*}[t]
\centering
\caption{Categories of Execution Violation}
\label{tab:Details of ESV}
\begin{tabular}{@{}r|r|rr|rr|rr@{}}
\toprule
\multicolumn{1}{c|}{Bug Category} &
  \multicolumn{1}{c|}{\begin{tabular}[c]{@{}c@{}}Execution\\ Violation\end{tabular}} &
  \multicolumn{2}{c|}{\begin{tabular}[c]{@{}c@{}}Unrecognized \\ Execution\end{tabular}} &
  \multicolumn{2}{c|}{\begin{tabular}[c]{@{}c@{}}Incomplete \\ Execution\end{tabular}} &
  \multicolumn{2}{c}{\begin{tabular}[c]{@{}c@{}}Inconsistent \\ Execution\end{tabular}} \\ \midrule
stack-out-of-bounds  & 30  & 8  & 26.67\% & 3  & 10.00\% & 19 & 63.33\%  \\
slab-out-of-bounds   & 38  & 4  & 10.53\% & 10 & 26.32\% & 24 & 63.16\%  \\
global-out-of-bounds & 9   & 0  & 0.00\%  & 3  & 33.33\% & 6  & 66.67\%  \\
invalid-free         & 11  & 4  & 36.36\% & 3  & 27.27\% & 4  & 36.36\%  \\
double-free          & 3   & 0  & 0.00\%  & 0  & 0.00\%  & 3  & 100.00\% \\
use-after-free       & 24  & 4  & 16.67\% & 11 & 45.83\% & 9  & 37.50\%  \\
null-ptr-def         & 24  & 2  & 8.33\%  & 2  & 8.33\%  & 20 & 83.33\%  \\ \midrule
Sum                  & 139 & 22 & 15.83\% & 32 & 23.02\% & 85 & 61.15\%  \\ \bottomrule
\end{tabular}
\end{table*}

% Please add the following required packages to your document preamble:
% \usepackage{booktabs}
% \usepackage{multirow}
\begin{table*}[t]
\centering
\caption{Performance}
\label{tab:Token}
\begin{tabular}{@{}r|r|rrr|rrr|c@{}}
\toprule
\multicolumn{1}{c|}{\multirow{2}{*}{Bug Category}} &
  \multicolumn{1}{c|}{\multirow{2}{*}{Total}} &
  \multicolumn{3}{c|}{Token} &
  \multicolumn{3}{c|}{ESV Token} &
  \multicolumn{1}{c}{\multirow{2}{*}{\begin{tabular}[c]{@{}c@{}}ESV Token/\\ Token\end{tabular}}} \\ \cmidrule(lr){3-5}\cmidrule(lr){6-8}
\multicolumn{1}{c|}{} &
  \multicolumn{1}{c|}{} &
  \multicolumn{1}{c}{Sum} &
  \multicolumn{1}{c}{Avg} &
  \multicolumn{1}{c|}{Max} &
  \multicolumn{1}{c}{Sum} &
  \multicolumn{1}{c}{Avg} &
  \multicolumn{1}{c|}{Max} &
  \multicolumn{1}{c}{} \\ \midrule
stack-out-of-bounds  & 28  & 641,540   & 22,912 & 68,771  & 213,927   & 7,640  & 43,854  & 33.35\% \\
slab-out-of-bounds   & 34  & 1,460,395 & 42,953 & 79,332  & 597,269   & 17,567 & 55,433  & 40.90\% \\
global-out-of-bounds & 11  & 445,852   & 40,532 & 80,468  & 217,636   & 19,785 & 64,637  & 48.81\% \\
invalid-free         & 16  & 479,002   & 29,938 & 90,040  & 246,462   & 15,404 & 66,534  & 51.45\% \\
double-free          & 4   & 129,372   & 32,343 & 56,640  & 71,301    & 17,825 & 33,327  & 55.11\% \\
use-after-free       & 33  & 765,089   & 23,185 & 56,426  & 210,606   & 6,382  & 47,125  & 27.53\% \\
null-ptr-def         & 44  & 1,440,530 & 32,739 & 58,714  & 912,173   & 20,731 & 56,357  & 63.32\% \\ \midrule
Sum                  & 170 & 5,361,780 & 31,540 & 90,040  & 2,469,374 & 14,526 & 66,534  & 46.06\% \\ \bottomrule
\end{tabular}
\end{table*}

\section{Evaluation}
\label{sec:evaluation}

In this section, we discuss the results of our evaluation. Specifically, we aim to answer the following research questions:
\squishlist
    \item \textbf{RQ1:} How accurate are the identified blamed functions of the prototype?~(\S\ref{sec:accuracy})
    \item \textbf{RQ2:} How effective is the Execution Specifications Verification process?~(\S\ref{sec:specification})
    \item \textbf{RQ3:} What is the overall performance of the Pseudo-code Execution and Execution Specifications Verification process?~(\S\ref{sec:performance})
\squishend

\noindent
\textbf{Configurations.}
In our experimental setup, we use the GPT-4 model with a 32k token limit~(\ie GPT-4-32k)~\cite{DBLP:journals/corr/abs-2303-08774} deployed in Microsoft Azure\cite{Azure}, to assess the effectiveness of our proposed method. 
We chose GPT-4-32k due to its current state-of-the-art performance across a wide range of language-based tasks. 
It represents the most sophisticated and capable option available to us for this study. 
We set the temperature parameter to zero to enforce deterministic behavior during generation. During the re-analysis because of the execution violation, we increase the temperature parameter by 0.2 each time until the temperature reaches 2.0~(\ie 10 times), which could generate more random results.

\noindent
\textbf{Dataset.}
Our evaluation is applied to bugs found in the Linux kernel, specifically those reported by syzbot~\cite{syzbot}. 
We focus on the Linux kernel for several reasons. 
First, it is among the most widely used open-source software projects in the world, thus attracting great interest from the research community. 
Second, the Linux kernel is a large and complex software system, which is representative and robust for evaluating the effectiveness of our proposed approach. 
Moreover, syzbot offers convenient access to both bug reports and their corresponding patches, which can be employed as ground truth in our evaluation.

Our evaluation specifically targets KASAN~(Kernel Address Sanitizer)~\cite{KASAN} bugs due to their impact on security. 
As of our last data collection on July 1, 2023, we extracted a total of 1140 fixed KASAN bugs. 
However, it is worth mentioning that not all of these bugs were suitable candidates for our evaluation. 
We applied a filtering process based on two primary reasons to narrow down our dataset:
\squishlist
    \item \textbf{Completeness of Call Trace:} 
    Our prototype is designed for bugs discovered through fuzzing techniques, which should, in principle, generate a complete call trace. 
    Bugs with incomplete call traces were thus excluded.
    \item \textbf{Reliability of Fix Commit/Patch:} 
    While syzbot reliably reports most of the bugs it finds, some older bugs~(those found before 2019) may have been initially reported by others. 
    These older bugs could have incorrect fix commits in syzbot. 
    Since we employ the fix commit/patch as ground truth in our evaluation, we excluded these older bugs from our dataset.
\squishend

Ultimately, as shown in \autoref{tab:Accuracy} we selected 170 fixed KASAN bugs from syzbot for our evaluation. 
These bugs meet all our requirements: they are reported by syzbot with a complete call trace and an accurate fix commit. 
The selected bugs include seven categories. 
For categories such as stack-out-of-bounds, global-out-of-bounds, invalid-free, double-free, and null-pointer-dereference, we included all valid bug reports from syzbot. 
In the case of slab-out-of-bounds and use-after-free, we randomly selected 34 out of 243 and 33 out of 634 valid bug reports from syzbot, respectively.

\subsection{Accuracy of Identified Blamed Functions}
\label{sec:accuracy}

\autoref{tab:Accuracy} shows the accuracy of our prototype in identifying the blamed functions for each bug category. 
We manually evaluated the correctness of the blamed functions identified by our prototype using the same standard in RETracer~\cite{DBLP:conf/icse/CuiPCFK16}. 
Specifically, an identified blamed function is considered correct if it meets one of the following conditions: (1) The patch is in the blamed function; (2) The patch is in a direct callee of the blamed function; (3) The patch is in a function that is related to the blamed function, \eg the same file or the same module. 
Those three scenarios are considered correct because those functions are likely to have the same developers for bug fixes.

Overall, we observe that our prototype achieves an overall accuracy of 81.18\% in identifying the blamed functions. 
The accuracy varies across different bug categories. 
For double-free bugs, our prototype achieves an accuracy of 50.00\%. 
This result is not considered statistically reliable due to the small number of bugs in the sample~(\ie only 4 bugs). 
For other bug categories, our prototype achieves an accuracy ranging from 72.73\% to 90.91\%.

We further conducted an in-depth manual analysis to understand the reasons for the wrongly blamed functions identified by our prototype. 
For all those wrong cases, we first confirmed that the underlying backward taint analysis was performed accurately. 
Then We categorize the reasons for these inaccuracies into three primary types:
\squishlist
  \item \textbf{High-level Fix:} 
  The patch for the bug is applied at a high level in the code base, implying that the modified function in the patch is not the one directly associated with the erroneous function. 
  So a backward taint analysis from the bug can not identify the correct function.
  \item \textbf{Source Code Issue:} 
  In some cases, our source code retrieval component fails to locate the appropriate source code, such as when a function is defined via macros.
  \item \textbf{Patch Not in Bug Call Trace:} 
  In some scenarios, the patched functions do not appear within the call trace leading to the bug. 
  For example, in memory-related bugs like slab-out-of-bounds, invalid-free, or use-after-free, the fix may involve either the memory allocation or release process rather than the direct bug call trace. 
\squishend

\subsection{Effectiveness of Execution Specifications Verification}
\label{sec:specification}

We further investigate those correct cases to understand the effectiveness of our \textbf{Execution Specifications Verification} process. The results are shown in \autoref{tab:Improvement of ESV}. 
We compare the results of our prototype with and without ESV. 
Specifically, we give the results of our prototype without ESV~(\ie PE), with ESV~(\ie PE+ESV), and the improvement~(\ie $\Delta$ESV). 
Overall, we observe that ESV improves the accuracy of our prototype by 52.94\%~(\ie from 28.24\% to 81.18\%). 
This result demonstrates the effectiveness of our \textbf{Execution Specifications Verification} process. 
For different bug categories, ESV improves the accuracy of our prototype by 45.45\% to 67.86\%. 
This result demonstrates that ESV is effective for different bug categories. 
The results show that when leveraging large language models for static analysis, verification is a very important part.

We then analyze the executions of our prototype for the correct cases improved by \textbf{Execution Specifications Verification} process~(\ie $\Delta$ESV). 
The results are shown in \autoref{tab:Rate of ESV}. 
Overall, there are 25.46\%~(\ie 139/546) execution violations among all executions of pseudo-code. 
There are 6.07 executions for each case on average and the maximum number of execution is 19. 
There are 1.54 execution violations for each case on average and the maximum number of execution violations is 4. 
For different bug categories, there are 1.14 to 2.00 execution violations for each case on average. 
The results demonstrate that not all executions need the \textbf{Execution Specifications Verification} process. 
However, any execution violation indicates that the final results are not correct.

\autoref{tab:Details of ESV} shows more details of execution violations. 
As we describe in \S\ref{sec:Verification}, we classify the execution violations into three types: (1) \textbf{Unrecognized Execution}, (2) \textbf{Incomplete Execution}, and (3) \textbf{Inconsistent Execution}. 
Overall, we observe that there are 15.83\%~(\ie 22/139) Unrecognized Execution, 23.02\%~(\ie 32/139) Incomplete Execution, and 61.15\%~(\ie 85/139) Inconsistent Execution. 
For different bug categories, there are up to 36.36\% Unrecognized Execution, up to 45.83\% Incomplete Execution, and 36.36\% to 83.33\% Inconsistent Execution. 
The results demonstrate that for all bug categories, the verification of Inconsistent Execution always plays a major role. 
And the verification of Unrecognized Execution and Incomplete Execution is also important for some bug categories, \eg invalid-free, use-after-free, and global-out-of-bounds.

\subsection{Performance}
\label{sec:performance}

\autoref{tab:Token} shows the performance of our prototype. 
We compare the total number of tokens~(\ie Token) and the total number of tokens for Execution Specifications Verification~(\ie ESV Token). 
Overall, we observe that the total number of tokens for \textbf{Execution Specifications Verification} is 46.06\%~(\ie 2,469,374/5,361,780) of the total number of tokens. 
This result demonstrates that an increase of 46.06\% in the total number of tokens brings an increase of 52.94\% in the accuracy of our prototype from 28.24\% to 81.18\%. 
For different bug categories, the total number of tokens for Execution Specifications Verification is 27.53\% to 63.32\% of the total number of tokens.

% \section{Case Study}
\section{Limitations and Future Work}
\label{sec: Limitations and Future Work}

\noindent
\textbf{Capabilities of Large Language Models in Simulating Complex Pseudo-code Execution for Static Analysis.}
\tool is based on the capabilities of LLMs in executing basic pseudo-code structures such as sequences, conditional branches, loops, and recursive functions.
However, we only demonstrate that the current most powerful large language model GPT-4-32k can handle them, especially the loops and recursive functions.
Therefore, it is necessary to have a thorough evaluation of the maximum complexity that other LLMs can handle, especially the number of loops and recursive calls in the pseudo-code. 
One possible future research direction is to systematically evaluate the capabilities of LLMs by increasingly complex pseudo-code. 
Besides, we only apply our method to backward taint analysis, leaving a wide range of other static analysis techniques unexamined. 
Another possible future research direction is to diversify the application of pseudo-code execution across various static analysis domains.

\noindent
\textbf{Automatic Generation of Pseudo-code and Execution Specifications.}
The pseudo-code is the bridge between the static analysis and LLMs in \tool.
People already attempted to prompt LLMs to generate or optimize prompts for downstream tasks~\cite{zhou2023large, yang2023large}.
It would be promising if we could prompt LLMs to automatically generate the pseudo-code for static analysis.
At the same time, since LLMs could help refactor the code~\cite{fan2023large}, another possible future research direction is to perform a round of refactoring on the pseudo-code by LLMs.
Besides, the execution specifications verification is a key step in \tool to mitigate the input conflicting hallucination of LLMs during the pseudo-code execution.
However, the current execution specifications are also generated by humans together with the pseudo-code.
Since currently we define three kinds of violations of execution specifications, it is possible to automatically generate those violations of execution specifications for the pseudo-code.

% \section{Threats to Validity}
\section{Conclusion}
\label{sec: Conclusion}
% \ziqiao{Limit the scope of our work. It makes me feel our tool is a general tool that can replace any static analysis. }
In conclusion, by integrating pseudo-code execution with LLMs, we have developed a novel method that significantly reduces the need for extensive human effort and expertise traditionally required in static analysis. 
This method not only simplifies the static analysis process but also addresses the challenges of hallucinations in LLMs by verifying the execution of the pseudo-code. 
Our prototype, specifically designed for triaging crashes through backward taint analysis, demonstrates accuracy and potential for practical application. 
The open-sourcing of our prototype furthers this contribution, encouraging collaborative development and innovation.

\clearpage
\bibliographystyle{ACM-Reference-Format}
\bibliography{reference}

\clearpage

\end{document}